\documentstyle[12pt]{article}
\newcommand{\beq}{\begin{equation}}
\newcommand{\eeq}{\end{equation}}
\newcommand{\be}{\begin{equation}}
\newcommand{\ee}{\end{equation}}
\newcommand{\bi}{\bibitem}

\def\Journal#1#2#3#4{{#1} {\bf #2}, #3 (#4)}

\def\NPB{{\em Nucl. Phys.} B}
\def\PLB{{\em Phys. Lett.}  B}
\def\PRL{\em Phys. Rev. Lett.}
\def\PRD{{\em Phys. Rev.} D}

\def\APJ{\em Astrophysical Journal }

\def\be{\begin{equation}}
\def\ee{\end{equation}}
\def\bea{\begin{eqnarray}}
\def\eea{\end{eqnarray}}
\def\bi{\bibitem}

\begin{document}
\begin{center}
{COSMOLOGY: THEORY AND OBSERVATIONS} \\
 A. D. DOLGOV  \\
{\it{Teoretisk Astrofysik Center\\
 Juliane Maries Vej 30, DK-2100, Copenhagen, Denmark
\footnote{Also: ITEP, Bol. Cheremushkinskaya 25, Moscow 113259, Russia.}
}}
\end{center}

\begin{abstract}
The comparison of the Standard Cosmological Model (SCM) with astronomical
observations, i.e. theory versus experiment, and with the Minimal Standard
Model (MSM) in particle physics, i.e. theory versus theory, is discussed.
The main issue of this talk is whether cosmology indicates new physics
beyond the standard $SU(3)\times SU(2)\times U(1)$ model with minimal particle
content. The answer to this question is strongly and definitely "YES". New,
yet unknown, physics exists and cosmology presents very weighty arguments
in its favor.
\end{abstract}

\section{Cosmological Parameters}

\subsection{Hubble Expansion Law }\label{subsec:hubble}
The cosmological stage is quite simple and is set by General Relativity
and the assumption (well confirmed by the data) of the homogeneous and
isotropic matter distribution on cosmological scales. The
Einstein equations for this very special but realistic case
were solved by Friedmann~\cite{aaf}, who found in particular that the universe
expands, i.e. distant objects run away in accordance with the law
\be
v= Hr
\label{vhr}
\ee
which was later observed by Hubble~\cite{eh}. Here $v$ is the velocity of a
distant cosmic object (a galaxy or a galactic cluster), $r$ is the the
distance to this object and the coefficient of proportionality $H$ is the well
known Hubble constant. (It is better to say instead the Hubble parameter
because, though $H$ is believed to be independent of space points,
its magnitude depends on time.) The proportionality law (\ref{vhr}) is quite
accurately established but the magnitude of the Hubble parameter is rather
poorly known. It is usually parameterized in the form
\be
H= 100\, h\,\, {\rm km/sec/Mpc}
\label{h}
\ee
where the dimensionless parameter $h$ is crudely bounded by $0.5<h<1$.
The recent data have a trend towards lower values. The average presented in
the review~\cite{prim} a few months ago is $h=0.6 \pm 0.1$. However the most
recent measurement~\cite{lt} by Hubble Space Telescope (HST) from brightest
cluster galaxies gives $h=0.89 \pm 0.1$.

\subsection{Cosmological Matter/Energy Density}\label{subsec:dens}

The matter (or better to say energy) density in the universe, $\rho$,
is characterized by the dimensionless parameter
\be
\Omega = {\rho \over \rho_c}
\label{omega}
\ee
where $\rho_c = 3H^2 m_{Pl}^2 /8\pi = 1.88\cdot 10^{-29} h^2\,\, {\rm g/cm^3}$
is the critical or closure density. It is usually said that if $\Omega \geq 1$
the universe is open (or spatially flat for the particular value
$\Omega =1$) and  will expand forever. If $\Omega < 1$ the universe is closed
and the expansion will later change to contraction and
the universe will re-collapse. These
statements about the ultimate universe fate are true only if vacuum energy
or, what is the same, cosmological constant is zero.

Theoretically most favorable value is $\Omega = 1$. It is the only value
which does not change with time in the course of the universe evolution.
Any initial deviation from unity would evolve with time as a power of
scale factor and thus would change by many orders of magnitude from initial
moment till present time. In inflationary scenario this initial value
is adjusted to 1 with exponentially good precision so that even after
the scale factor changes in the course of the normal Friedmann expansion
more than by 30 orders of magnitude, the deviation from unity remains
very small. Inflationary cosmology
predicts $\Omega = 1 \pm 10^{-4}$ on the present day horizon scale.
The indicated here possible deviations from unity are not related to the
initial non-perfect adjustment of $\Omega$ to unity but
could be induced by the local density
perturbations. In principle open or closed universe (with $\Omega \neq 1$)
might be compatible with inflationary models but at the expense of an
unnatural fine-tuning.

There are different contributions into $\Omega$ coming from different forms
of matter, $\Omega = \sum_j \Omega_j$. Directly
visible matter gives a minor fraction into it:
\be
\Omega_{vis} \leq 0.003 h^{-1}
\label{omegalum}
\ee
Presumably all visible matter consists of usual baryonic staff which we
observe in our surroundings. A large part of baryons is most probably
invisible.
The cosmic baryon budget is analyzed in ref.~\cite{fhp} where it is
concluded that
\be
0.007 \geq \Omega_B \leq 0.04
\label{budg}
\ee
There is an independent way to find the density of cosmic baryons based on
primordial nucleosynthesis (see below). This method is sensitive to the total
baryonic number density and gives:
\be
\Omega_B^{NS} = (0.7 - 3) \cdot 10^{-2} h^{-2} = (0.7 -12) \%
\label{omegans}
\ee
If the Hubble constant is at the lower end of the permitted interval,
$h=0.5$, the fraction of invisible baryons is rather large,
$\Omega_{vis} /\Omega_B \leq 0.21$, while for the large $H$, $h=1$,
the fraction of invisible baryons is much smaller,
$\Omega_{vis} /\Omega_B \leq 0.43$, and it is easier to conceal them in the
universe. Still the question remains where are all these invisible baryons.
Cold baryonic gas could be observed by absorption lines in quasars. Hot gas
would produce an unacceptable distortion of the cosmic microwave
background or emit too much X-rays. As was recently shown in ref.~\cite{cr},
$\Omega_B > 0.1$ would result in production of too many X-ray bright
clusters. Dust would be seen by observation in
infrared range.

A good hiding place for invisible baryons are
compact objects such as white dwarfs
or black holes. A large fraction of white dwarfs is difficult to explain
with the standard theory of stellar evolution. Too many black holes which
were produced as result of stellar collapse are excluded because of
accompanying
enrichment of the interstellar space by heavier elements. To overcome this
difficulty the black holes should be primordial but, if so, they could
equally well, or even most probably, consist of non-baryonic matter. A direct
search of compact objects by gravitational micro-lensing in our Galaxy and
galactic halo, the so called MACHO's, is in process. More than a hundred
of such objects have been already observed (for a recent review see
e.g. ref.~\cite{bs}). There are indication that masses of these MACHO's are
quite large, $m> 0.3 m_\odot$, where $m_\odot$ is the Solar mass. That heavy
compact objects, if they are made of normal matter, should be luminous,
practically normal stars. The absence of light from these objects is quite
mysterious. If e.g they are white dwarfs then their age must be bigger
than 18 Gyr which is not compatible with the measured value of the Hubble
parameter if cosmological constant is zero (see below).

Anyhow the baryonic mass fraction of the universe is quite small. One can
safely conclude that $\Omega_B < 0.1$ so that the cosmic mass/energy
density is dominated by a non-baryonic (dark) matter. This conclusion is
supported both by direct observations and by the theory of large scale
structure formation.

It is known for already quite a long time that masses of galaxies and their
clusters are not concentrated inside the luminous central part but
are spread over a much larger distances where no matter is directly seen.
This is observed by the velocity of gas around the luminous centers up to
distances almost 10 times larger than the galactic radius. The velocity
$v(r)$ remains constant with increasing distance.
It means that the mass density is non-zero even outside
the galactic radius. It falls down as $\rho_m \sim 1/r^2$ and the total
mass inside radius $r$ is not a constant as one would normally expect but
rises as $m(r) \sim r$. A large sample of these flat rotational curves can
be found in ref.~\cite{sp}. The analysis of the mass-to-light ratio found
from galactic rotational curves up to the scales of 100 kpc permits to
conclude that $\Omega_{DM} \geq 0.1$. The same analysis made for galactic
clusters at the scales of several Mpc leads to the conclusion that
$\Omega_{DM} \geq 0.2$. Different methods and results of the determination
of $\Omega$ are reviewed in refs.~\cite{wf,prim}. Different estimates vary
in the region 0.2-0.4, while larger values are not excluded. Possibly safe
bounds are
\be
\Omega_m \geq 0.3\, ,
\label{omegam1}
\ee
\be
\Omega_\Lambda \leq 0.7\, .
\label{omegam}
\ee
Here $\Omega_m$ describes the mass fraction of the normal matter, both baryonic
and non-baryonic, while $\Omega_\Lambda$ describes contribution of vacuum
energy density into total cosmological energy/mass density. The flat universe
with $\Omega_{tot} = \Omega_m +\Omega_\Lambda =1$ is not excluded.

\subsection{Universe Age}\label{subsec:age}

If one knows the present-day values of the Hubble parameter $H_0$ and
contributions to $\Omega$ from different forms of matter one can calculate
the universe age:
\be
t_U = \int^1_0 {dx \over
\left( 1 - \Omega_{tot} + \Omega_{m}x^{-1} +\Omega_{rel}x^{-2}
+\Omega_{\Lambda}x^{2} \right) ^{1/2}}
\label{tu}
\ee
It is assumed normally $\Omega_{rel}=0$, $\Omega_{\Lambda}=0$ and the universe
age is approximately given by the expression
\be
t_u\approx { 9.8 \,h^{-1}\, {\rm Gyr} \over 1+\sqrt \Omega /2 }
\label{tu1}
\ee
Nuclear chronology gives universe age in the interval 11-17 Gyr. The estimate
of the universe age from the ages of old globular clusters before spring 1997
gave \cite{rh} $t_U = 14\pm 2$ Gyr. Recent observations made by astronomical
satellite Hipparchos showed that the distances to globular clusters are
systematically 5-10\% larger so that the stars in fact are brighter and
younger. With these new data the universe age found from globular clusters
becomes somewhat smaller:
\be
t_U = 12 \pm 2 \,{\rm Gyr}
\label{tu3}
\ee
The smaller universe age and possibly low Hubble constant, $h=0.6$,
are compatible with flat universe $\Omega_{tot} =1$ without cosmological
constant. So possibly the age crisis which existed for larger $t_U$ and
high $H_0$ is over. Still one should keep in mind that the data are not
yet conclusive.

\section{ A Few Problems}

Though the simple isotropic homogeneous cosmology quite well describes the
universe at large, there are a few disturbing points for which no natural
explanation is known.
\begin{enumerate}
\item
{\it $\Omega$-conspiracy.} Inflationary cosmology naturally predicts that
$\Omega$ is very close to 1 but it remains unclear why contributions to
$\Omega$ from different forms of matter are of the same order of magnitude.
As we mentioned above, baryonic contribution is at the level of 1\%. The
dynamically measured $\Omega$ is somewhere in the range 0.3-1. The theory of
structure formation requests at least two different forms of dark matter:
hot dark matter (HDM) and cold dark matter (CDM) with
$\Omega_{HDM} = 0.2-0.3$ and $\Omega_{CDM} \approx 0.7$ respectively.
The contributions
of different forms of matter may differ by many orders of magnitude and their
approximate equality looks quite mysterious.
\item
{\it Cosmological constant.} Observations do not exclude that cosmological
constant is non-vanishing. The contribution of the corresponding vacuum energy
into $\Omega$ can be as large as $\Omega_{\Lambda} \sim 0.7$. Moreover the
theory of structure formation with just one form of dark matter (CDM) favors
non-zero $\Lambda$ of this order of magnitude. In the course of the universe
expansion the energy densities of all normal forms of matter drops down with
the scale factor either as $1/a^3$ for non-relativistic matter or $1/a^4$ for
relativistic matter while the contribution from $\Lambda$ stays constant. It is
another mystery why exactly today these contributions happened
to have similar magnitudes.
Astronomers mostly prefer vanishing cosmological constant following Einstein
who considered introduction of $\Lambda$ as the greatest blunder of his life.
However there is no consensus with regard to the value of $\Lambda$ and
for example Le Maitre and Eddington believed that cosmological constant
could be non-zero.
\item
{\it Vacuum energy.} Though astronomers discuss if $\Lambda$ is identically
zero
or essential on cosmological scales, with corresponding energy density of the
order of $\rho_{vac} \sim \rho_c \approx 10^{-47} {\rm GeV}^4$, from the point
of view of particle physicists it should be many orders of magnitude, 50-100,
larger than the astronomical upper bound. There are several contributions into
vacuum energy each of which is by far bigger than the permitted upper bound
(for the recent discussion see e.g. refs. \cite{ad1,ad4}). In particular there
is
the quite well known contribution to vacuum energy from quark and gluon
condensates which give
$\rho_{vac} \approx 10^{45} \rho_c$. There must exist some other
contribution to $\rho_{vac}$ which cancels out the quark-gluon contribution
with the fantastic precision better than one part per $10^{45}$. This is
one of the most striking problems of the modern fundamental physics. To my mind
the best solution of this problem is the so called adjustment mechanism when
there exists a new field which is unstable in the De Sitter space-time and
which vacuum condensate induced by the curvature of space-time automatically
cancels down any initial vacuum energy \cite{ad2,ad1}. Though no satisfactory
model of this kind yet has been found, one can conclude that they possibly
possess a common feature that the vacuum energy is not completely compensated
but only down to the terms of the order of $\rho_c \sim m_{Pl}^2/t^2$. If
this is indeed realized then the non-compensated part of vacuum energy always
gives a contribution to $\Omega$ of order of unity.
\end{enumerate}

\section{Cosmic Microwave Background}

Cosmic microwave background (CMB) is one of the pillars supporting Big Bang
cosmology. It has a perfect Planckian spectrum \cite{cobe} with the temperature
$T = 2.728 \pm 0.001$ K. It is very uniform. There is a noticeable dipole
component in the angular distribution of the radiation which corresponds
to $\Delta T = 3.35 $ mK. This dipole is most probably related to our motion
with respect to the preferable cosmic frame where microwave radiation is at
rest. The measured value of the quadrupole, which describes an inherent
anisotropy of CMB is about $10 \mu K$. Angular distribution of CMB is measured
in terms of spherical harmonics up to $l\sim 1000$. It could be one of the
most powerful tools for determination of cosmological parameters,
$\Omega$, $H$, $\Lambda$. The position and the magnitude of the so called
acoustic peaks in the angular distribution depends upon the values of these
parameters. The planned cosmic missions MAP and PLANCK will
be able to determine these parameters with the accuracy of 10\% and 1\%
respectively. This is a considerable improvement in comparison with the present
day 50-100\% accuracy. Even with the present date data the maximum in the
angular distribution around $l=100-200$ is already observed. It is an
interesting possibility if the position of this maximum corresponds to the
observed 100 Mpc scale in the large scale structure of the universe
\cite{silk}.

\section{History of the Universe}

I. {\bf Beginning} is unknown. Several possibilities are considered:
\begin{enumerate}
\item
Creation from nothing.
\item
Perpetually oscillating universe.
\item
Eternal chaotic inflation.
\item
Pre-big-bang string cosmology.
\end{enumerate}
Discussion of these possibilities is outside the scope and volume of the this
talk. At the present stage there is no way to distinguish between these
possibilities but what is quite certain a description of the universe
creation definitely needs new physics.

\noindent
II. {\bf Inflationary stage.} During this stage the universe expanded
exponentially, $a\sim \exp (H_I t)$, with a constant (time-independent) $H_I$.
This simple assumption perfectly solved all initial value problems of the
Friedmann cosmology~\cite{guth} if $H_I t > 60$. During this stage the universe
is dominated by the vacuum-like matter with the equation of state
\be
\rho + p = 0
\label{rhop}
\ee
Because of the covariant conservation of the energy-momentum tensor,
\be
\dot \rho = -3H (\rho + p)
\label{dotrho}
\ee
it follows that the energy density remains
constant in the course of expansion. This surprising result means that all
the matter in the universe, all universe mass might be created from
microscopically small initial piece of matter which approximately
satisfied equation (\ref{rhop}).

One can ask if inflation is a necessity or the universe can reach the
present stage without inflationary period. The no-go theorem is not proved
but there is no known way to create our suitable for life universe without
inflation. In this sense inflation is an experimental fact. We do not see
any other possibility to solve simultaneously the following problems:
\begin{enumerate}
\item
{\it Flatness.} The present day value of parameter $\Omega$ is rather close
to 1. It means that during primordial nucleosynthesis when light elements
($^2H$, $^3He$, $^4He$, and $^7Li$) were synthesized in good agreement with
observations, $\Omega$ should be close to 1 with the precision of $10^{-15}$.
At the "initial" Planck epoch the adjustment must be much better, about
$10^{-60}$. No other model except inflation naturally gives such a fine-tuning.
\item
{\it Homogeneity, isotropy, horizon}. The universe looks the same in any
direction from us though in the old Friedmann cosmology different parts of the
universe could not be in causal contact. Such similarity means that expansion
regime must be different in the past so that the regions which seem to be
out of contact in fact came from the same microscopically small piece of
space with the same initial conditions. Exponential expansion could
easily achieve that.
\item
{\it Initial push.} We see that the universe expands but the origin of the
initial push which created this expansion remained mysterious in the Friedmann
cosmology. Inflation naturally explains that because the particular equation
of state (\ref{rhop}) gives rise to gravitational {\it repulsion}
(anti-gravity).
It is possible only for infinitely large objects for which the Gauss theorem
cannot be applied. For such objects the source of gravity is $(\rho + 3p)$
and with $p= -\rho$ it induces anti-gravity for normal positively definite
energy density.
\item
{\it Density fluctuations.} Though the universe is very smooth on large scales,
at smaller scales it has a very rich structure, stars, galaxies, galactic
clusters, etc (?). For their creation some initially small density
perturbations
must exist, which later on rose up due to gravitational instability and evolved
into the observed structure. Of course quantum or thermal fluctuations existed
in the early universe but their wave length was by far too small to be
cosmologically interesting. Inflation could amplify initially small quantum
fluctuations (maybe a little too much) and to stretch their initially
microscopic wave length up to astronomical scales. This solves the problem
of generation of density perturbations. The perturbations have a specific
spectrum (flat spectrum) and this can be tested by measurement of angular
distribution of CMB.
\end{enumerate}
An unnatural feature of inflationary scenarios is that
the scalar field (inflaton) which drives inflation should have a tiny coupling
to other field as well as a tiny constant of self-interaction,
$\lambda<10^{-14}$. Such a field does not exist in the minimal standard model
and for the realization of inflation a new scalar field is necessary.

\noindent
III. {\bf End of Inflation}
A very important period in the history of the universe is the transition from
inflationary stage to the "normal" Friedmann stage. During inflation the
inflaton field, $\phi$, remains (almost) constant. This ensures validity of
equation of state (\ref{rhop}). After some period the temporal evolution of
$\phi$ becomes non-negligible. Generically $\phi (t)$ begins to oscillate
around
minimum of its potential energy and, as any oscillating field, starts to
produce elementary particles which thermalize and form primeval cosmic plasma.
A simplified theory of universe heating was considered in
refs.~\cite{dl,reh,reh1}. The particle production was treated perturbatively
and it
was shown that the transition of energy from inflaton field proceeded
slowly so that the temperature of heating was rather small. The
non-perturbative approach to heating was put forward in refs.~\cite{dk,bt}.
It was noticed there that the inflaton decay in principle could be
parametrically excited and, if this is the case, the process of heating would
proceed much faster and efficiently than in the naive theory. However as
was argued in ref.~\cite{dk} the universe expansion and re-scattering of the
produced quanta destroy the resonance and it became ineffective. Recently
the problem was re-addressed in ref.~\cite{kls} and in a large number of
subsequent papers where is was shown that parametric resonance might be excited
and the heating possibly proceeded quite fast. In particular, in such a case
the decay of inflaton might produce super-heavy intermediate bosons of grand
unification and this would permit baryogenesis at GUT scale.

\noindent
IV. {\bf Baryogenesis.}
The universe in our neighborhood consists predominantly of matter.
Antiprotons and positrons observed in cosmic rays most probably have
a secondary origin. It is normally assumed that all visible universe is
built of the same kind of matter as our Galaxy. It is a very interesting
question if there may exist a considerable amount of (primordial) antimatter
in the universe. And, if "yes", how far away could be these antimatter
domains. This problem was addressed recently in refs.~\cite{der,kolb} where
it was argued that the smoothness of CMB does not permit antimatter domains
to be closer than a Gygaparsec. However some rather exotic models of
baryogenesis~\cite{ds} may produce regions of abundant antimatter not so far
away.

The asymmetry between matter and antimatter in the universe may be
either explained by
asymmetric initial conditions ensuring dominance of matter over antimatter
or by a dynamical generation of the asymmetry from initially symmetric
or even arbitrary initial state. This generation of charge (or baryonic)
asymmetry is called baryogenesis. Baryogenesis can be realized if the
following three natural conditions are fulfilled~\cite{ads}:
\begin{enumerate}
\item
Non-conservation of baryonic charge.
\item
Breaking of C- and CP-invariance.
\item
Deviation from thermal equilibrium.
\end{enumerate}
If inflation existed, then baryogenesis is a necessity. It can be shown that
with any initial conditions conservation of baryonic charge is not compatible
with inflation~\cite{ad3}.

A natural frameworks for baryogenesis are presented by grand unification
theories (GUT). In all such theories baryons are not conserved. Deviation
from thermal equilibrium induced by the universe expansion is quite
significant.
Though nothing is known about CP-violation at GUT scale, it is easy to
believe that once CP is broken in low energy physics it is also broken
at high energies.  With possible efficient universe heating after inflation
by excitation of parametric resonance in inflaton decay GUT baryogenesis may
still be a viable possibility.

Another beautiful possibility is baryogenesis on electroweak scale~\cite{krs}.
Electroweak interactions are known to break baryonic charge
conservation~\cite{gth} by non-perturbative effects related to quantum chiral
anomaly and quantum barrier penetration. It is also well known that
both C and CP are broken by electroweak interactions. Deviation from
thermal equilibrium is more difficult to realize at electroweak scale than
at GUT scale but even this is possible if electroweak phase transition is first
order. It may be so for sufficiently light Higgs bosons. Thus all conditions
for
baryogenesis exist even in the minimal standard model of particle physics.
However all attempts to find a satisfactory scenario which may explain
the observed baryon-to-photon ratio, $n_B/n_\gamma \approx 3\times 10^{-10}$,
have failed. The minimal electroweak model gives roughly 5-10 orders of
magnitude smaller result. Possibly a low energy supersymmetric extension of
MSM is in a better shape and could explain the observed baryon asymmetry.
One essential point about general applicability of electroweak approach to
baryogenesis is that it is usually assumed that particular field configurations
which realize baryonic charge non-conservation (so called
sphalerons~\cite{sphal}) are abundant in primeval plasma. It is assumed that
their number density at high temperatures is the equilibrium one. Strictly
speaking it is not known. There is no reliable analytic way to estimate the
production rate of sphalerons. The only available possibility now are lattice
calculations. Unfortunately they are not accurate enough and the results of
different authors~\cite{amb,arn} disagree.

Anyhow, baryogenesis is a necessary feature of modern cosmology and it could
generate observable asymmetry only in an extension of MSM. These could be
either a low energy supersymmetric extension or high energy SUSY or GUT but
definitely new fields and new interactions absent in MSM are necessary.

\noindent
V. {\bf Primordial Nucleosynthesis}
The discussed above phenomena in the early universe lay to some extend in
{\it terra incognita} where our knowledge of fundamental physics is
non-complete or even absent. Primordial nucleosynthesis takes place at
low energies or temperatures, from $\sim 1$ MeV down to 100 keV, in the
time interval from 1 sec to 200 sec. Everything at this stage is quite
well known from direct nuclear physics experiments and the predictions of the
theory of light element formation in the early universe are quite robust.
During this stage the following light elements were produced:
$^4He$ (25\% by mass), $^2H$ ($2\cdot 10^{-5} - 2\cdot 10^{-4}$ by number,
relative to hydrogen) and similar amount of $^3He$, and
$^7 Li$ ( a few $\times 10^{-10}$ by number). The predictions of the theory are
in a good agreement with observations, though they span 10 orders of magnitude
in relative abundances.

Primordial nucleosynthesis serves as a good "cleaner" for different exotic
possibilities in particle physics. It permits to put an upper bound on the
number of extra neutrino species or other particles which were abundant during
nucleosynthesis. The present day limit on extra neutrino species
or other abundant at nucleosynthesis particles
is $\delta N_\nu <1$ (for the recent review see
e.g. ref.~\cite{olive}). It permits in particular
to put an upper bound on the possible
mass of tau-neutrino, $m_{\nu_\tau} < 1 $ MeV (the recent most accurate
calculations of the influence of massive $\nu_\tau$ and the list of
appropriate references can be found in paper~\cite{dhs}).

Two years ago the accuracy of the bounds derived from primordial
nucleosynthesis was believed to be considerably better than now but recent
conflicting data on primordial deuterium abundance created some confusion.
Several groups~\cite{tfb,rh1,tbk,swc,dt1,wcl}
have reported measurements of the deuterium abundance
in Lyman-limit absorption line systems with red-shifts $0.48<z<3.5$ on the
line of sight to quasars; these are believed to give essentially the
primordial value. Surprisingly some groups have claimed a high value,
$ D/H \approx 2\cdot 10^{-4}$ on the basis of ground-based data taken
with the Keck telescope, but this result is now thought to be due to various
errors \cite{tbk} and the best value available from two ``clean" systems
is $3\cdot 10^{-5}$ \cite{dt1}. However, Webb et al \cite{wcl} report a high
deuterium abundance,  $ D/H \approx 2\cdot 10^{-4}$, in an apparently clean
system with $z = 0.7$ observed with the Hubble Space Telescope, as well as
a low one in another system with $z = 0.5$, raising the possibility that
there might be real spatial variations in primordial $D/H$.
If the effect is indeed  real (which it is perhaps too early to judge),
its significance is difficult to overestimate. It would
strongly change our approach to primordial nucleosynthesis and possibly to the
physics of the early universe. In the literature there are two explanations
of the possible spatial variation of deuterium: either by non-homogeneous
baryogenesis~\cite{jf,cos} or by large and spatially varying leptonic chemical
potentials~\cite{dp}. In the first case one should either expect too large
variation of the CMB temperature~\cite{cos} or variation of deuterium on very
small scales~\cite{jf} which is possibly forbidden by observation. In the case
of large and varying lepton asymmetry the model~\cite{dp} predicts
a very large mass fraction of primordial $^4He$ (up to 50-60\%) in
deuterium-rich regions. Surprisingly this is not excluded by the data.

To conclude, primordial nucleosynthesis is quite well described by the
standard well known physics but if spatial variation of primordial deuterium
indeed exists, a new physics seems necessary for an explanation of the
phenomenon.

\noindent
VI. {\bf Structure Formation.}
The early universe was quite homogeneous as a result of inflation; it is well
confirmed by the isotropy of CMB. At the present day we observe evolved
large scale structure. By assumption this structure was formed as a result of
gravitational instability from initially small density perturbations. This is
the basic point of the theory of large scale structure formation and it seems
that (almost?) everybody agrees with it.

Initial density perturbations most probably were generated during inflationary
stage. Their spectrum is an important input in calculations of the structure
formation. It is usually assumed that the original spectrum is flat (without
any particular scale) as follows from simple inflationary models. However
other forms of spectrum are possible and this may introduce a lot of freedom
into theory, though this possibility is not very popular.

Another important input parameter is the chemical composition of the
universe. Structure formation is very much suppressed in the universe
which consists only of the usual matter, that is of baryons and electrons.
When the temperature of primeval plasma was above 3000 K the matter was
ionized and structure formation was suppressed due to a large radiation
pressure.
The structure could be formed only after hydrogen recombination at or below
red-shift $z = 10^3$. In such a case there is too little time for formation
of evolved structures. So a new form (or several different forms) of matter
which is (are) not strongly coupled to radiation is (are)
very desirable (or maybe necessary). To this end we also need new physics.
Particle physics has several possible candidates on the role dark matter
particles but it is unknown which one indeed plays this role. Possibly a
comparison of calculations of structure formation with detailed
future astronomical data will permit to deduce some properties of dark
matter and to single out the right particles.

\section{Conclusion}

It seems quite well established that the universe during its evolution
went through epochs of
inflation, baryogenesis, nucleosynthesis, and structure formation. Without
them it would be simply impossible to create the observed world. Surprisingly
physical processes during practically all these epochs demand an extension
of the Minimal Standard Model. Moreover there are some cosmological puzzles
which may need new physics for their resolution.
Let us compare Standard Cosmological Model with Minimal Standard Model of
particle physics summarizing discussion of the previous sections:
\begin{enumerate}
\item
{\it Inflation.} Does not exist in MSM. Needs a new scalar field with mass
$m< 10^{13} $ GeV and small coupling of self-interaction, $\lambda < 10^{-14}$.
\item
{\it Baryogenesis.} Possible in MSM but too weak. For successful realization
needs supersymmetric extension of MSM (though there may be some serious
problems), GUT's or some other new physics at high energies.
\item
{\it Nucleosynthesis.} In a good agreement with MSM. A possible trouble is
a spatial variation of primordial deuterium. If confirmed, it may indicate
on a very serious problem in both MSM and SCM.
\item
{\it Structure formation.} MSM does not produce primordial fluctuations on
astronomical scales. To this end one needs inflation or topological defects,
though the latter seem to be excluded now.
\item
{\it Dark matter.} It is seen by its gravitational action and is necessary for
structure formation. The only possible candidate in MSM is massive neutrino
(or neutrinos), but the observed structure cannot be described with this
form of dark matter only. New particles are necessary: lightest supersymmetric
particle (if R-parity is conserved and it is stable), axion, majoron, shadow
world particles, etc. One of the most interesting challenges is to find
what is the nature of dark matter particles constituting 90\% or more of the
total mass of the universe.
\item
{\it Vacuum energy.}  The discrepancy between theoretical expectations and
observations is 50-100 orders of magnitude (!). There is no satisfactory
solution to this problem even with any kind of new physics.
\item
{\it  Conspiracy of $\Omega_j$.} Close values of $\Omega_j$ for different
forms of matter. There is no explanation for that in MSM. Possibly it can be
found in unified theories.

\end{enumerate}

There are some more more unsolved problems which are not related to
cosmology on large scales but still present a serious theoretical challenge.
It is not known if they can be solved in the frameworks of the usual physics
but of course these attempts should be done first. Among them are
$\gamma$-bursters, magnetic fields in galaxies, cosmic rays of extremely
high energies, non-luminous heavy MACHO's and possibly some more.

Thus even though MSM is in
very good agreement with practically all direct experiments we still can be
sure that New Physics exists. With the new generation of telescopes,
with new more sophisticated technology we should expect that new discoveries
and more mysteries are on the way.

\section*{Acknowledgments}This work was supported by Danmarks
Grundforskningsfond through its funding of the Theoretical Astrophysical
Center (TAC).


\end{document}